\RequirePackage{lineno}
\documentclass[aps,prl,twocolumn,superscriptaddress]{revtex4}

\usepackage{graphicx}
\usepackage{savesym}
\savesymbol{tablenum}
\usepackage{siunitx}
\usepackage{amsmath}
\usepackage{amssymb}

\usepackage[version=4]{mhchem}
\DeclareMathOperator{\EMD}{EMD}

\begin{document}
\title{Machine-learning prediction of infrared spectra of interstellar polycyclic aromatic hydrocarbons}

\author{P\'eter Kov\'acs}
\affiliation{Institute of Materials Chemistry, TU Wien, 1060 Vienna, Austria}

\author{Xiaosi Zhu}
\affiliation{Laboratory for Relativistic Astrophysics, Department of Physics, Guangxi University, 530004 Nanning, China}

\author{Jes\'us Carrete}
\affiliation{Institute of Materials Chemistry, TU Wien, 1060 Vienna, Austria}

\author{Georg K. H. Madsen}
\affiliation{Institute of Materials Chemistry, TU Wien, 1060 Vienna, Austria}

\author{Zhao Wang}
\email{zw@gxu.edu.cn}
\affiliation{Laboratory for Relativistic Astrophysics, Department of Physics, Guangxi University, 530004 Nanning, China}
\affiliation{Institute of Materials Chemistry, TU Wien, 1060 Vienna, Austria}

\begin{abstract}

We design and train a neural network (NN) model to efficiently predict the infrared spectra of interstellar polycyclic aromatic hydrocarbons (PAHs) with a computational cost many orders of magnitude lower than what a first-principles calculation would demand. The input to the NN is based on the Morgan fingerprints extracted from the skeletal formulas of the molecules and does not require precise geometrical information such as interatomic distances. The model shows excellent predictive skill for out-of-sample inputs, making it suitable for improving the mixture models currently used for understanding the chemical composition and evolution of the interstellar medium. We also identify the constraints to its applicability caused by the limited diversity of the training data and estimate the prediction errors using a ensemble of NNs trained on subsets of the data. With help from other machine-learning methods like random forests, we dissect the role of different chemical features in this prediction. The power of these topological descriptors is demonstrated by the limited effect of including detailed geometrical information in the form of Coulomb matrix eigenvalues.

\end{abstract}

\maketitle
\section{Introduction}
Polycyclic aromatic hydrocarbons (PAHs) are among the most widely studied organic compounds in the fields of astronomy \citep{Herbst2009}, chemistry \citep{Zhang2014}, biology and environmental science \citep{Moorthy2015,Ravindra2008}. As some of the most abundant molecules in the universe \citep{Snow1995}, PAHs are understood to play an essential role in the evolution of the interstellar medium (ISM) \citep{Tielens2008,Hardegree2014,McGuire2018,Qihn2018}. They are also thought to have acted as elemental building blocks of complex organic molecules related to the origin of life \citep{Ehrenfreund2000}. Since the infrared (IR) spectrum of a molecule contains valuable information of the molecular bonding configuration \citep{Neubrech2017,Meier2003}, IR spectroscopy has become an indispensable tool in many observatory projects such as the Stratospheric Observatory for Infrared Astronomy and the Spitzer Space Telescope \citep{Young2012,Deming2020}.

Due to the complex structure-property relationship, identifying PAHs from their IR spectra is anything but straightforward. Since IR activity is related to changes in the molecular dipole moment, a good characterization requires knowledge of both the dynamics of the atomic nuclei, specifically in the form of a set of normal modes, and of the electronic charge distribution. In many cases, the best option available at a reasonable computational cost is density functional theory (DFT). Unfortunately, the number of possible existing PAH species in the ISM is so vast that a brute-force application of DFT is unlikely to be successful in interpreting experimentally measured IR spectra from mixtures of arbitrary molecules \citep{Croiset2016,Andrews2015,Shannon2018}. Indeed, while the ``unidentified'' infrared emission (UIE) features dominating the mid-IR spectra of a wide variety of interstellar sources has been linked to PAHs \citep{Allamandola1999,Maltseva2015,Bouwman2019} the exact chemical species responsible for UIE are still under debate \citep{Kwok2011,Li2012,Kwok2013}. Therefore, developing efficient approaches to the prediction of IR spectra of interstellar PAH remains an important goal with a view to the accurate identification of the UIE band carriers among other sources.

Recently, the rapid development of machine-learning (ML) methods has opened new and reliable ways of investigating molecular structure-property relationships \citep{Butler2018,Marquez2018,Ghosh2019,Gastegger2017}. However, vibrational spectra are a challenging property for any ML method as they cannot be explained in terms of global composition or local bonding, but depend on hybridizations involving many atoms. In the present study we aim at developing a neural-network (NN) based accelerated model to predict the IR spectra of PAHs using just their skeletal formula as input. Such formulas encode the topology of the molecule without reference to the exact coordinates of the atoms and, despite their abstraction of the geometric details, are central in any discussion of the structure of organic molecules and exhibit a large amount of predictive power. They thus provide an ideal starting point for the kind of accelerated model that we develop here and do not require computationally intensive electronic structure calculations to determine optimized geometries.

We present an efficient, data-driven approach to the prediction of the IR spectra of PAHs that combines a NN and inputs extracted from the NASA Ames PAH IR spectroscopic database. The potential of NNs to predict IR spectra of organic molecules was explored more than 20 years ago \citep{Weigel1996,Selzer2000}. However, the discriminatory power of these pioneering attempts was not particularly convincing. Moreover, recent developments in deep learning have led to large improvements in the effectiveness of NN. The results of a NN depend intricately on the descriptors and penalty function used for training. In the present paper we demonstrate that it is possible to obtain good predictive power from a multilayer NN trained on Morgan descriptors that represented skeletal formulas. We discuss the success of these descriptors in detail and show how including the Coulomb matrix \citep{Rupp2012,Schutt2014}, which has otherwise been very successful for encoding molecular structure, does not improve the predictive power of the model.

We furthermore train random forests (RFs) on the same data. Generally speaking, RFs have lower quantitative predictive skill than well trained NNs. However, they can be easier to interpret. For instance, RFs have an intrinsic metric for the importance of each feature that can be computed simply by reverting all the choices based on that feature. Moreover, they are naturally resistant to overfitting and work well with correlated inputs. In this study we use NNs to provide quantitative predictions and RFs to take a closer look at the effect of adding and removing information from the input.

\section{Methods}

\subsection{Dataset}
The NASA Ames Research Center has assembled computed and experimental IR spectra of PAHs into a public database, available since 2010 \citep{Bauschlicher2010}. This database comprises more than three thousand spectra, and has undergone two major updates \citep{Boersma2014,Bauschlicher2018}. It includes PAH IR spectra obtained using DFT, as well as a number of spectra from experimental measurements. This database is an important tool for determining the IR spectra of PAHs in order to develop and test hypotheses regarding astronomical PAHs \citep{Allamandola1989,Draine2007,Peeters2011,Tielens2008}.

For this work, we take the structures and IR spectra of all PAHs in the computational dataset version 3.00. As detailed in the next subsection, we use topological descriptors, so we discard those cases in which several geometries exist that are compatible with the same topology. Such cases include, but are not limited to, topologically equivalent structures with different charge states. That leaves us with $2670$ molecules from the $3129$ in the database. We then turn the set of discrete lines of the IR spectrum into a histogram with a bin width of \SI{21.39}{\centi\meter^{-1}} determined using Knuth's Bayesian rule \citep{Knuth2006}. Each histogram consists of $252$ bins covering the range from \SI{6.95}{\centi\meter^{-1}} to \SI{5376}{\centi\meter^{-1}}. Bins beyond the $176$th (i.e., beyond \SI{3751}{\centi\meter^{-1}}) are discarded since there is a single compound in the whole database contributing to this region of the spectrum with a single (and possibly spurious) peak. We then split each of the truncated histograms into a low-frequency and a high-frequency part. The splitting is done at $106$th bin (\SI{2253}{\centi\meter^{-1}}) because it lies in the middle of a gap without contributions from any compound in the database. Frequencies above this cutoff typically correspond to localized vibrations involving hydrogen atoms. Finally, each of the two sub-histograms is normalized with the obvious exception of histograms composed entirely of zeros. At the end of this round of preprocessing, the IR spectrum of each compound in the database is represented by two vectors, one for the low-frequency part of the histogram and the other for the high-frequency part, and the components of each add up to one. Those vectors are the targets for the prediction.

\subsection{Loss function}
A key piece of a good machine-learning model is a suitable loss function, i.e., a target to be minimized during the training process. A common way to build such a function is to introduce a notion of distance between output values and then sum the distances between the known and predicted values of the output over the training set. In the context of the current application, each of those values comes in the form of an array representing a normalized histogram. Therefore, it is of critical importance to define a sensible idea of distance between two histograms that takes into account the nature of the elements in those arrays. Among the requirements for that distance is that a slight misprediction of the position of a line should contribute less to the distance than a significantly larger error in that prediction. General-purpose distances like the Euclidean norm of the difference between histograms do not fulfill this criterion, since they do not take the distance between bins into account. Therefore we opt for a more specialized function, in particular a version of the earth mover's distance (EMD) \citep{Monge1781, Dobrushin1970}. Introduced in 1781 in the context of literal transport of dirt between two sites and now known to be a special case of the Wasserstein metric, the EMD measures the minimal cost of transforming a histogram into another when the cost of moving a unit of mass from bin $i$ to bin $j$ is set to a fixed non-negative value $c_{ij}$. We specifically make $c_{ij}$ proportional to the distance between the center of the bins, $\left\vert i - j\right\vert$. With this choice, if ones takes the spectrum of a molecule and introduces a random perturbation in all frequencies of the order of the bin width, the distance between the two histograms will be rather small, and in particular much smaller than the distance to another arbitrary molecule. In contrast, big errors in the placement of lines will increase the distance much more significantly. Moreover, this particular choice of costs allows for a simple and efficient implementation of the EMD. Let $\mathbf{a}=\left(a_i\right)_{i=1}^N$ and $\mathbf{b}=\left(b_i\right)_{i=1}^N$ be two normalized histograms with the same set of bins, and $\left(A_i\right)_{i=1}^N$ and $\left(B_i\right)_{i=1}^N$ the corresponding cumulative histograms, with $A_i = \sum_{j=1}^i a_j$ and a similar expression for $B_i$. The distance between the histograms is computed as:

\begin{equation}
  \EMD\left(\mathbf{a}, \mathbf{b}\right) = \sum\limits_i \left\vert A_i - B_i\right\vert = \sum\limits_i \left\vert \sum\limits_{j\le i}\left(a_j - b_j\right)\right\vert.
  \label{eqn:EMD}
\end{equation}

\noindent There is a clear connection with other measures of differences between distributions like the Kolmogorov-Smirnov statistic \citep{Smirnov1944}, in which the distance is the maximum value of $\left\vert A_i - B_i\right\vert$.

\noindent In addition to its role in building the loss function, we also employ the EMD to evaluate the quality of a prediction and to quantify the similarity between two spectra.

\begin{figure}[ht]
    \centering
    \includegraphics[width=0.3\textwidth]{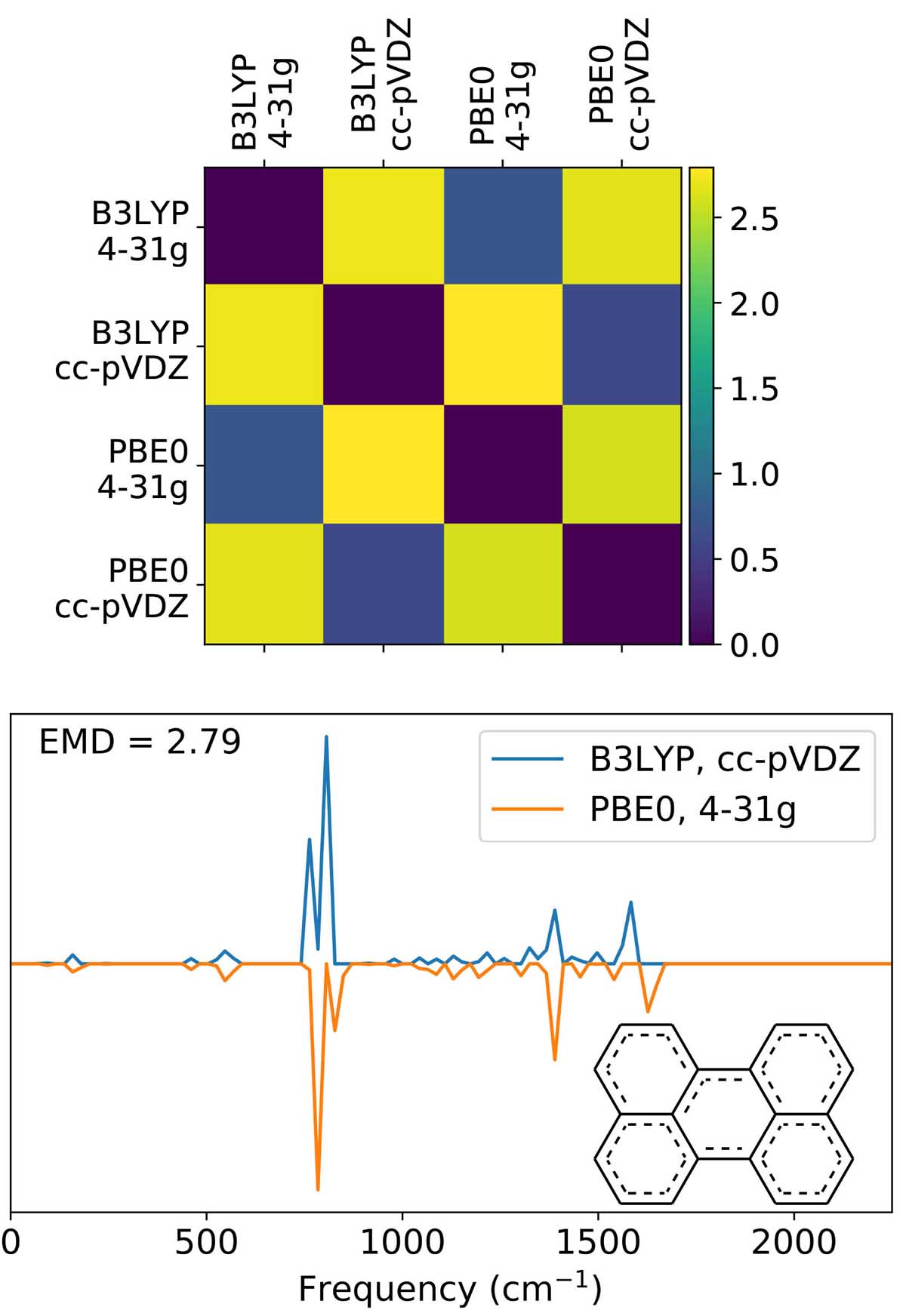}
\caption{Top: EMD between the DFT-calculated IR spectra obtained using different functionals and basis sets for perylene. The database contains the spectrum from a B3LYP, 4-31g calculation. Bottom: Direct comparison of the most dissimilar IR spectra predicted by DFT calculations with different functionals and basis sets in the case of perylene.}
  \label{fig:cross_method_emd}
\end{figure}

To illustrate the distribution of EMD values, the IR spectra of perylene is calculated with two different hybrid functionals and basis sets using the NWChem software package \citep{nwchem}. The spectra are then scaled according to the prescription for version 3.0 of the database \citep{Bauschlicher2018}, which splits the peaks in three different regions and scales the frequencies for these regions individually to get better agreement with experimental results. Fig.~\ref{fig:cross_method_emd} shows the EMD values between these calculated spectra. As expected, the (B3LYP, 4-31g) calculation matches the data included in the database. A change in the functional introduces only small differences, whereas changing the basis set causes larger EMD values between the calculated spectra. The worst agreement can be found between the (B3LYP, 4-31g) and (PBE0, cc-pVDZ) calculations, as illustrated in the bottom half of Fig~\ref{fig:cross_method_emd}. 
The EMD in that case is $2.79$, which we will use as a reference value for ``good'' predictions since they are comparable to the variance among DFT calculations. 

\subsection{Descriptors}
\begin{figure}[ht]
    \centering
  \includegraphics[width=.45\textwidth]{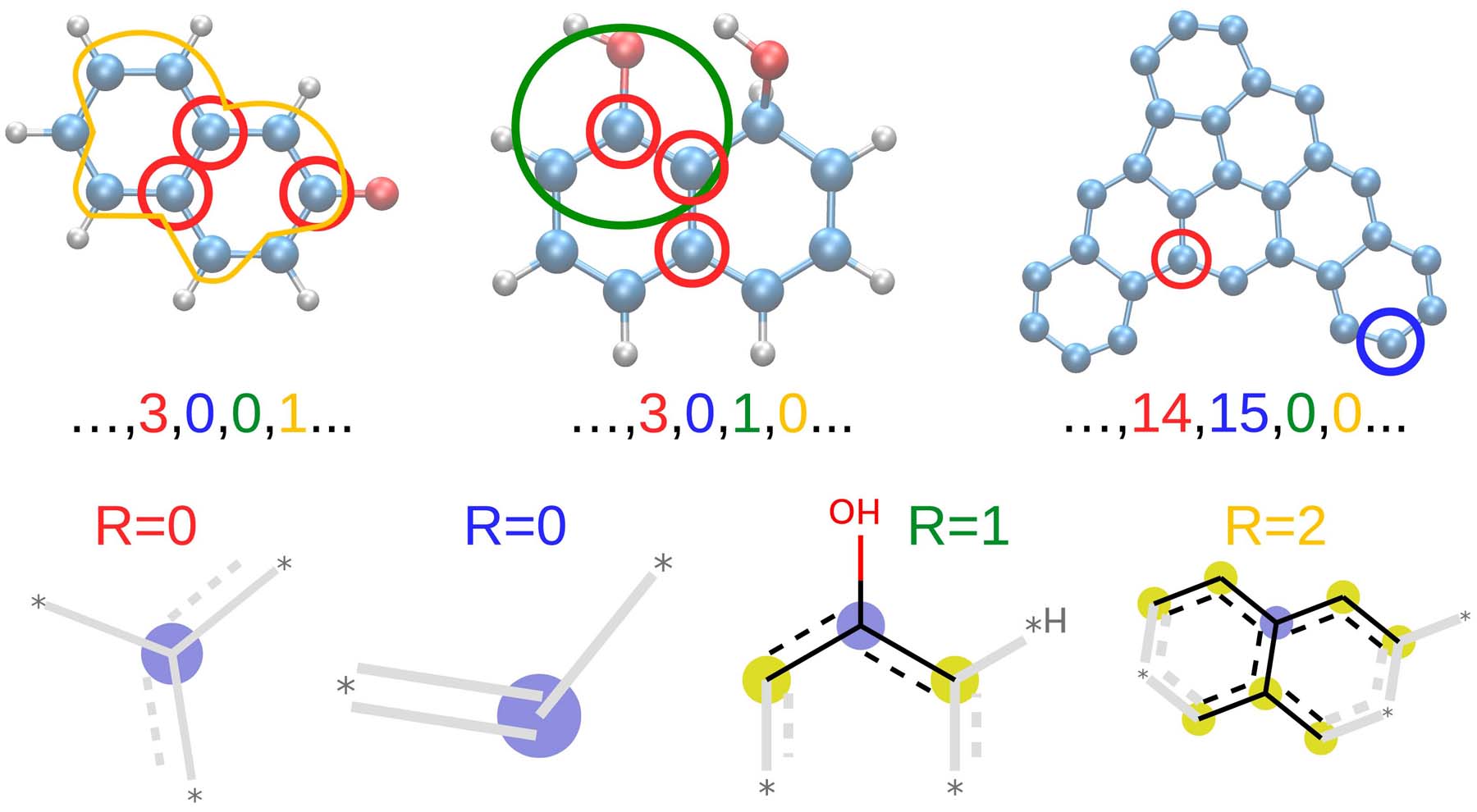}
  \caption{Illustration of how the topological descriptors are built based on the presence of different molecular fragments. Example molecules are shown on top, with marks showing the positions of the specific fragments depicted below. The middle line shows the corresponding region of the generated counting fingerprints. The red and blue fragments are generated by the first iteration of the fingerprint generation, so they only contain information about their base atom. In the current case, the red items represent carbon atoms with $3$ non-hydrogen neighbours and no hydrogens connected to them, while the blue items represent carbon atoms with $2$ non-hydrogen bonds and, likewise, with no hydrogen neighbours. The green and yellow circles show fragments generated by the 2nd and 3rd iterations respectively. During training we use more than $9200$ unique fragments generated by up to $11$ iterations of the algorithm.}
    \label{fig:descriptors}
\end{figure}

We focus on topological fingerprints as our primary descriptors of the skeletal formula. Often used for substructure and similarity searching, such fingerprints express whether a molecular graph contains particular subgraphs and how many copies of those subgraphs it contains. We specifically use the implementation of extended connectivity fingerprints (ECFPs) \citep{Rogers2010} in RDKit \citep{rdkit}. These fingerprints are calculated using a modified version of the Morgan algorithm \citep{Morgan1965}, originally designed to create a canonical numbering scheme for atoms in molecules. Each non-hydrogen atom is initially assigned a $32$-bit integer identifier derived from the properties used in the Daylight atomic invariants rule \citep{Weininger1989}. The algorithm then proceeds for a predefined number of iterations, replacing that identifier with the hash of an array formed by the identifiers of the atom and its first neighbors listed in a deterministic order. In Fig.~\ref{fig:descriptors} four examples of substructures are shown, generated by 0 (red, blue), 1 (green) and 2 (yellow) iterations. The results of all iterations are put together and the occurrences of each substructure counted to create the final fingerprint. To be able to detect large substructures of potential relevance for the low-frequency portions of the spectrum, we perform $11$ iterations of the algorithm.

The structures are extracted from the database in the form of a set of atomic coordinates, which are then converted to the SDF format \citep{sdf} using Open Babel \citep{Boyle2011} and finally to a simplified molecular-input line-entry system (SMILES) string from which the aforementioned descriptors are extracted. Since the XYZ-to-SDF conversion involves the use of bond-detection heuristics not guaranteed to work in every case, the conversion fails for $18$ compounds and those are dropped at this stage.

As mentioned in the previous section, topologically equivalent structures with different charge states were removed from the dataset. The remaining charged molecules with unique descriptors make up around $7\%$ of the dataset. The average EMD for these molecules ($5.3$) is around $2.\mathbf{9}$ times larger than that for the neutral ones ($1.8$). We conclude that charged molecules are not accurately described by the SMILES in our processing pipeline. To check if this has an impact on the results, we also train NNs only on neutral molecules. However, the average EMD remains the same and the only significant difference was a thinner tail in the EMD histogram. We therefore keep those charged molecules.

The topological fingerprints cannot encode geometric information, and we aim at developing a model that does not rely on such information. To assess whether this choice influences the result we also train models based on the eigenvalues of the Coulomb matrix \citep{Rupp2012}.

\subsection{Machine-learning models}
\subsubsection{Neural networks}
NNs are one of the most powerful families of ML techniques in use today \citep{Bishop1996}, and also one of the most widespread, partly because of the existence of high-performance implementations for both CPUs and GPUs. An NN is a graph consisting of layers of nodes, or neurons. In the cases of interest for this discussion, each node produces an output based on a linear combination of the nodes from the previous layer plus a constant. Specifically, the output from a neuron is computed as

\begin{equation}
Y = f\left(\sum_{i=1}^N w_i X_i+b\right),
\end{equation}

\noindent where $X_i$ is the output of the $i$-th neuron in the previous layer, $w_i$ is the connection weight to the current neuron, $b$ is the bias and $f$ is the activation function, responsible for the non-linearity in the network. During training, all the weights and biases of the neural network are fitted so as to minimize a target penalty, which in the present case is the EMD, described by Eq.~\eqref{eqn:EMD}.

Due to their many and diverse applications, different NN architectures have been devised, such as convolutional NNs (where all neurons in a layer share their weights, but their sparse connections to the previous layer are displaced in a systematic way) or recurrent neural networks (where the output of the NN is fed to it again as an input). However, in our case we do not need to capture the type of features those architectures were designed for, so we opt for an archetypical fully connected multi-layer NN, where each neuron is connected to every node of the previous and the next layers. Aside from the input and the output layers, our network has four hidden layers with $1500$, $1000$, $850$ and $600$ neurons, respectively. As activation functions we use rectified linear units (ReLUs) \citep{Lecun2015} with an extra linear layer and absolute value function before the output.

The input dataset is randomly split in training, test and validation subsets containing $70\%$, $15\%$ and $15\%$ of the data, respectively. The inputs to the NN are the descriptors defined above, after removing all elements that did not appear in the training set. The feature vectors so constructed contain $9231 \pm 15$ elements. The weights of the NN are initialized using the Glorot algorithm \citep{Glorot2010} and all the biases are initialized to zero. The training is then carried out with an Adam optimizer \citep{ADAM} with the EMD as the target, until the validation error fails to decrease for $50$ consecutive epochs. The model is implemented, trained and evaluated using TensorFlow \citep{tensorflow2015-whitepaper}. The results presented are calculated with an ensemble of $40$ individual neural networks for whose training the training/validation/test split was performed in $40$ different ways, always according to the proportions quoted above.

\subsubsection{Random forests}
The second ML technique that we use is RFs \citep{random_forests}. Each RF regressor (or classifier, as the case might be) consists in an ensemble of classification trees, each of them trained on a random subset of the observations (a technique known as ``bagging'') and performing splits at each level of the tree based on a random subset of the variables. The result of the RF regression for a new structure is obtained by running the set of descriptors down each tree, obtaining the corresponding individual predictions, and averaging them. Since the prediction of each tree so built is always a value from the training set, a RF regressor of this type has a strong centralizing tendency \citep{PhysRevX.4.011019}. 

We use the implementation of RF in scikit-learn \citep{scikit-learn}. Our forests contain $1000$ trees each. A tree stops growing when leaf nodes contain just one element or when the depth (i.e., the number of splits) equals $15$. A maximum of four features are considered when looking for an optimal split. The splitting criterion is the minimization of the mean square error instead of the EMD. The RFs are mainly used as an interpretative tool and an EMD splitting criterion resulted in too long training times.

\section{Results and discussion}
\label{sec:results}

In the following we present the results of our NN and RF models.

\subsection{NN Performance Metrics}
\begin{figure}[ht]
\centering
\includegraphics[width=.45\textwidth]{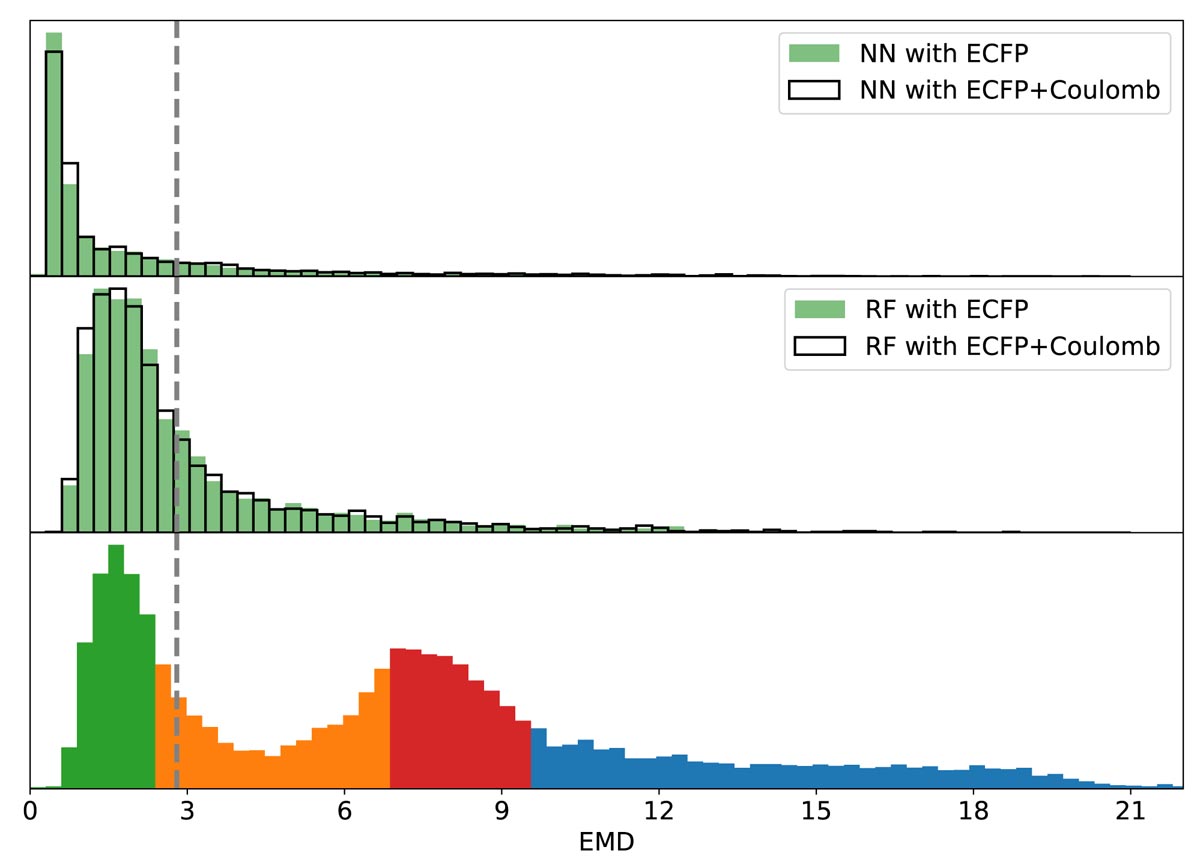}
\caption{Distribution of EMD values between each individual NN (top) and RF (middle) predictions and the database spectra. The green bars give the results for models trained using only the topological fingerprints and the empty bars the results for models trained also using the ten largest eigenvalues of the Coulomb matrix. The bottom panel shows the EMDs between random pairs of structures from the database. As a reference, the $25\%$, $50\%$ and $75\%$ percentiles of the EMDs between random pairs are shown as changes in color in the bottom panel. The grey vertical line shows the reference 2.8 EMD from Fig. \ref{fig:cross_method_emd}. All EMDs shown here correspond to the low-frequency parts of the spectra.}
\label{fig:histograms_NN}
\end{figure}

\begin{figure}[ht]
    \centering
    \includegraphics[width=0.45\textwidth]{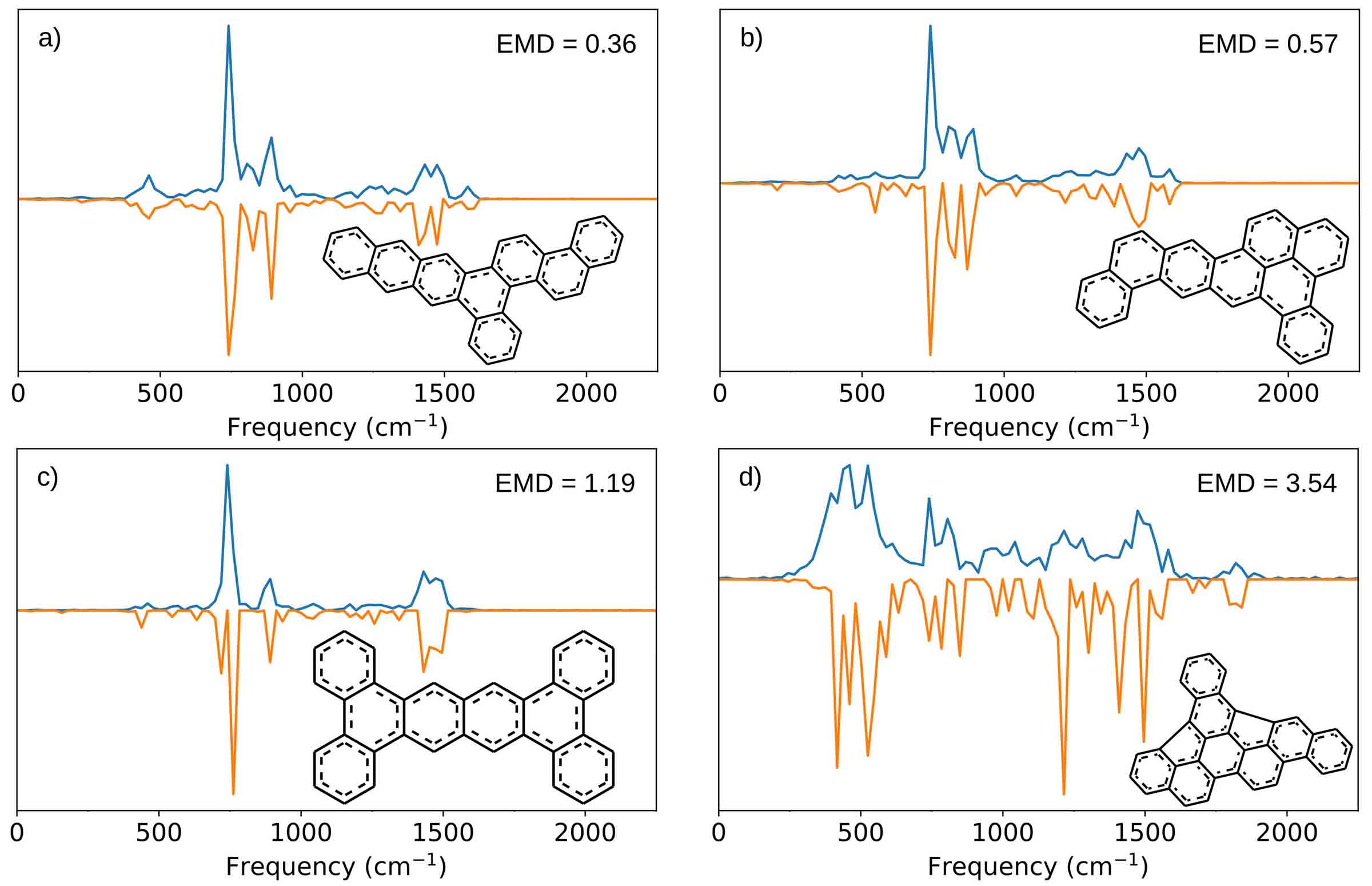}
  \caption{Comparisons between the database result (blue lines) and the NN prediction (orange lines) for four compounds in the test set drawn at random from the regions between (a) $0\%$ and $25\%$, (b) $25\%$ and $50\%$, (c) $50\%$ and $75\%$, and  (d) $75\%$ and $100\%$ of the distribution of EMDs. In other words, the four panels provide examples of what a prediction looks like for different levels of quality, from good to poor. All histograms shown here correspond to the low-frequency parts of the spectra.}
  \label{fig:illustration_NN}
\end{figure}

As described in the previous section, we train the NN architecture separately for the low- and high-frequency parts of the spectrum. Figs.~\ref{fig:histograms_NN} and \ref{fig:illustration_NN} provide, respectively, a quantitative and a qualitative window into the performance of the NN for the low-frequency part. This is often labeled the fingerprint region \citep{fingerprint_region}, is used to identify the molecule, and comprises most of the mass of the histograms. More specifically, Fig.~\ref{fig:histograms_NN} (top panel) shows how the model for the low-frequency part performs, by way of the distribution of the EMD between each database record of the test set and the corresponding prediction. Most of the EMD values are found well below the $2.8$ reference value extracted from DFT calculations on perylene with different parameters [Fig.~\ref{fig:cross_method_emd}].

A reasonable criterion for what values of the EMD can be considered as good is the discriminatory power, i.e., to be useful, a predicted spectrum for a molecule A should be significantly closer to the ``real'' spectrum of A than the spectrum of some other molecule B. Therefore, Fig.~\ref{fig:histograms_NN} also shows, in the bottom panel, the distribution of EMDs between pairs of structures in the database. The distribution is color coded according to the $25\%$, $50\%$ and $75\%$ percentiles. Most of the NN predicted spectra yield EMDs well below those marks, with $73\%$, $92\%$ and $95.7\%$ below the first, second and third quartiles, respectively, indicating a good predictive skill of the model. Another baseline is the average EMD of 5.79 between random samples of the database, which could indicate if the model has any predictive power at all.

An interesting feature of the baseline histogram in Fig.~\ref{fig:histograms_NN}(bottom) is its bimodal structure, with a clear divisory line around an EMD of $4.5$. Both of the peaks are well populated, with $\sim 20\%$ of the compound pairs in the central region of each. One possible explanation of this behavior could be that the database contains groups of molecules with relatively small intragroup EMDs, thus creating the first peak in Fig.~\ref{fig:histograms_NN}(bottom), and significantly larger intergroup EMDs, forming the second peak and the long tail that contains a further $\sim 25\%$ of the compound pairs. We put this hypothesis to the test using a clustering algorithm, specifically $k$-medoids \citep{elements_of_sl} because it allows us to use a custom metric for clustering, which we set to the low-frequency EMD depicted in the histogram. We select the optimum number of clusters as that maximizing the average silhouette score \citep{silhouette}, i.e., the average over all structures of $\left(b-a\right)/\max\left(a,b\right)$, where $a$ is the mean intracluster distance for that particular structure and $b$ is the distance from the structure to the nearest cluster other than its own. This optimum number turns out to be two; the clusters contain $67\%$ and $33\%$ of the structures, and their median intracluster EMDs are $2.2$ and $7.3$, respectively. Therefore, the initial hypothesis is false: as a matter of fact, the database consists of a core of closely related compounds (the first cluster) and a second cluster of more loosely similar ones. Each of the peaks in the baseline histogram contains the intracluster distances of one of these two clusters, and the tail of the distribution comes mostly from intercluster distances.

An additional indicator of the good performance of the NN model is the fact that this bimodal structure is absent from the top panel of Fig.~\ref{fig:histograms_NN}. This goes to show that the model does not act as a mere nearest-neighbor interpolant, looking for similar molecules whose spectrum to copy, but is actually able to pick up different structural features from each molecule and build an accurate prediction based on them.

As a more qualitative illustration, Fig.~\ref{fig:illustration_NN} shows four example spectra to provide the reader with an idea of what can be considered a good or a poor prediction in the context of our model. The four structures are chosen at random from each of the quartiles of the NN EMD distribution. The quartiles for the NN results are $0.49$, $0.82$, $2.54$ and $23$. This means that $75\%$ of the predicted spectra have an $\text{EMD}<2.54$, which is comparable to the EMD between DFT predictions obtained with different DFT parameterizations [Fig.~\ref{fig:cross_method_emd}]. Fig.~\ref{fig:illustration_NN} also illustrates how even relatively large EMDs are qualitatively informative, which underlines the suitability of the EMDs as a tailored distance metric and the modified Morgan fingerprints as a molecular descriptor.

The NN for the high frequencies yields a median EMD of $0.15$. While this value is remarkably small in the context of Fig.~\ref{fig:histograms_NN}, the median EMD between the high-frequency parts of two random histograms is only $0.33$. In fact, as will be seen in more detail below, the high-frequency parts of the spectra have a narrow unimodal distribution and contain far less detail that can or needs to be predicted. We trained a binary classifier to try and predict which molecular structures have a high-frequency part in their histograms at all. After trying both NNs and RFs for this task, we find that it is easy to build a classifier with perfect precision and almost perfect recall, that is, free of false positives and with a single false negative. The reason for this unusually high level of skill can be analyzed by looking at the importance of each feature in the RF model or by systematically pruning the input features in the case of the NN. Interestingly, in both cases it is revealed that nearly perfect classification can be obtained by using just a single feature, specifically the fingerprint bit shown as blue in Fig.~\ref{fig:descriptors}, representing an unsaturated carbon atom on the edge of an aromatic ring. In our dataset, the molecules containing multiple instances of the mentioned fragment have no hydrogen atoms. This points to C-H bonds as responsible for the localized, high-frequency vibrations, in agreement with physicochemical intuition. The finding provides an example of how ML techniques can replicate domain knowledge and, in particular, well known ``rules of thumb'', without any specific guidance from specialists.

\subsection{Feature Importance}
\begin{figure}[ht]
\begin{center}
  \includegraphics[width=0.45\textwidth]{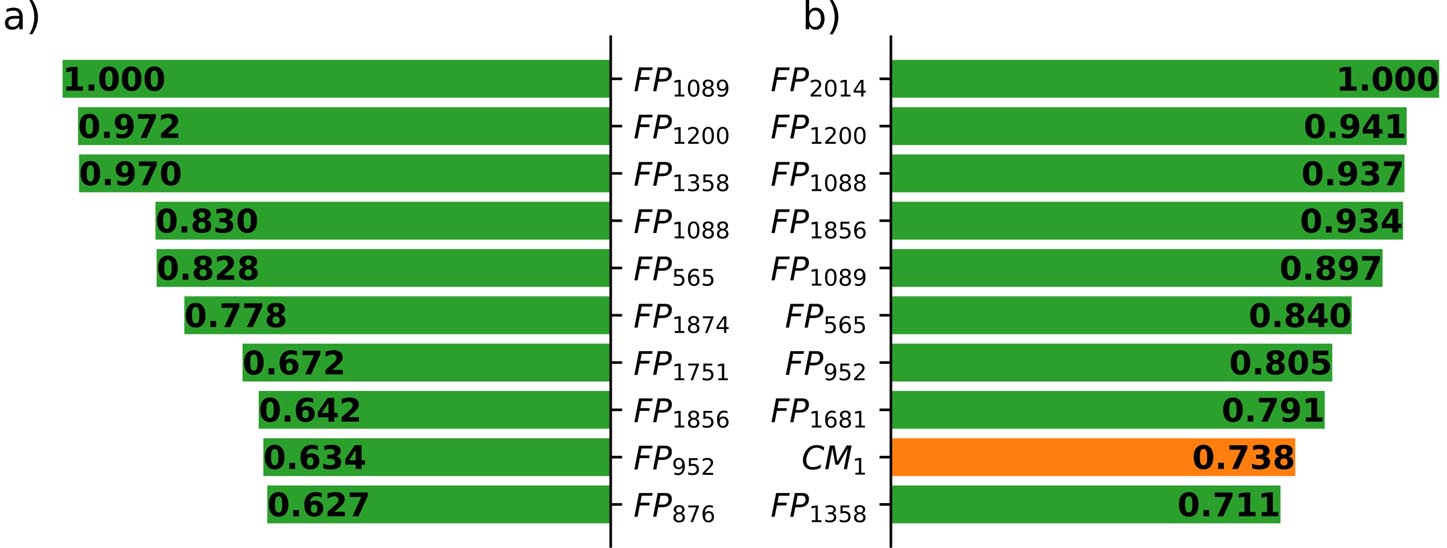}
  \caption{Most important features according to the random-forest models for the low-frequency part, when only the topological fingerprints are included in the input (left-hand-side panel) and when those are supplemented with the eigenvalues of the Coulomb matrix (right-hand-side panel). Green bars denote topological features, orange bars represent eigenvalues of the Coulomb matrix. Importances have been renormalized so that the first feature in each list has an importance of unity.}
  \label{fig:importance_rf}
  \end{center}
\end{figure}

Our next step consists in training a RF model based on the same dataset as the NN. As expected from the discussion in the previous section, the predictive power of the RFs is lower, partly because of the flexibility of the model and partly because the RF was trained to minimize the mean square error and not the EMD. This can be seen in the center panel of Fig.~\ref{fig:histograms_NN}, which shows the distribution of EMD in the test set. On the other hand, an advantage of RFs is the intrinsic feature importance metric they provide. In the left panel of Fig.~\ref{fig:importance_rf} we show the ten most important features of the low-frequency RF models trained on the topological fingerprints. The values in each list have been renormalized to assign an importance of one to the most important feature.

\begin{figure}[ht]
    \centering
  \includegraphics[width=.25\textwidth]{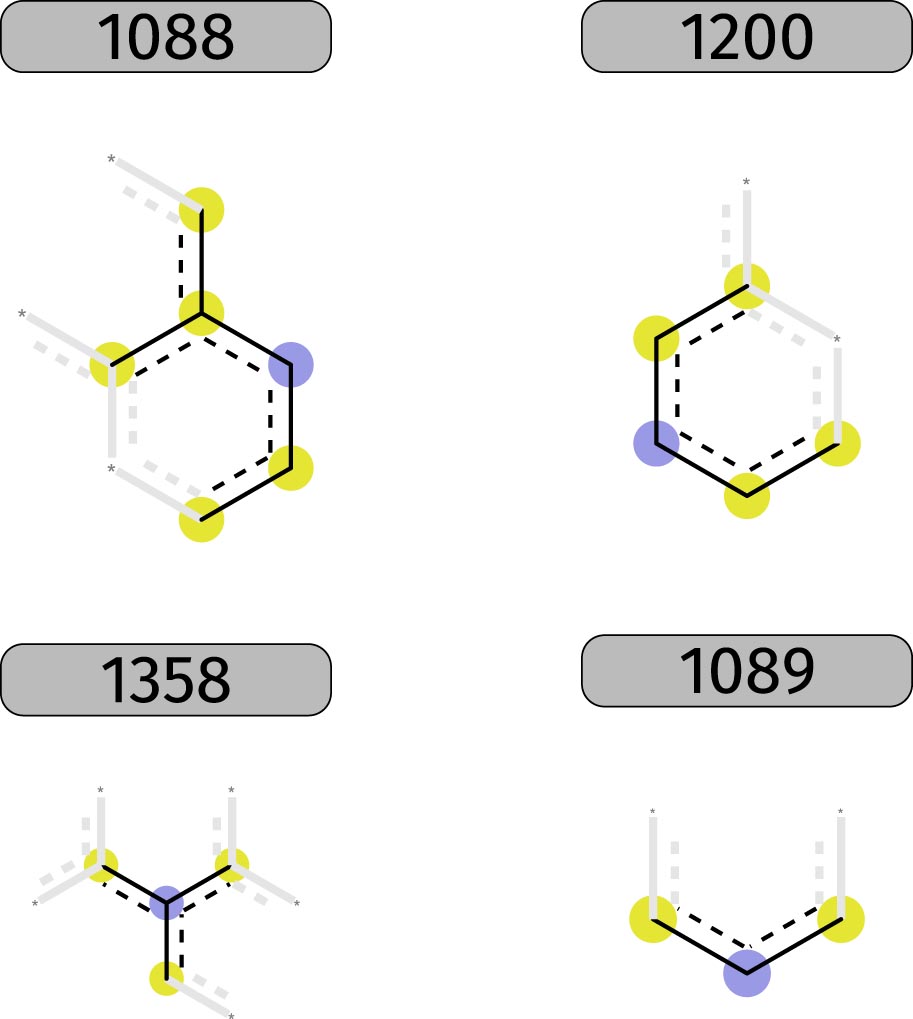}
  \caption{Most important molecular fragments and their unique identifiers for the low frequency prediction according to the random forest model.}
  \label{fig:fragments}
\end{figure}

A remarkable result is that four features are important for both low- and high-frequency predictions (not shown): $1088$ and $1089$, $1200$ and $1358$. The substructures that those features represent are depicted in Fig.~\ref{fig:fragments}. Their complexity reveals that ML models of vibrational behavior must consider sequences of many bonds to achieve good predictive skill. For the low frequencies this is intuitively obvious, since those normal modes arise from the hybridization of many individual vibrations and involve many atoms.

\subsection{Coulomb Matrix}

We then test whether the predictive power can be improved by adding information to the input that the topological fingerprints cannot encode, namely descriptors of the molecular geometry. A priori there are situations where IR spectra depend on their stereochemistry, for instance if the effect of non-bonded interactions between atoms far away in the molecular graph causes large changes in the vibrational frequencies of the structure. It is clear that nothing in the topological descriptors can directly address those situations. However, the real question is whether the connection between topology and geometry is strong enough for PAHs in interstellar space that the former can be used as a proxy for the latter.

To answer this question, the histograms on the top and center panels of Fig.~\ref{fig:histograms_NN} show the results of NN and RF regression models based on topological information only (filled) and models obtained when the topological descriptors are supplemented with the ten largest eigenvalues of the Coulomb matrix (unfilled contours). Comparing each pair of histograms reveals very little improvement in model performance coming from the addition of geometry, especially in comparison with the large differences introduced by switching the underlying model or the loss function. The lower importance of the Coulomb matrix eigenvalues compared to the fingerprints might be due to the loss of information inherent in using a few eigenvalues of a matrix with $4n-6$ independent elements that encode all structural information of an $n$-atomic molecule, but also point to part of the geometric structure being predictable from the topology itself, as expected.

The inclusion of the Coulomb matrix eigenvalues does have a discernible effect on the structure of the RFs. This is evidenced by a comparison between the left- and right-hand-side panels of Fig.~\ref{fig:importance_rf}, which list the ten most important features in the models built without and with that information, respectively. However, the four features discussed above remain among the most ten most important and only one eigenvalue enters this group. It does thus not seem probable that including more structural information would improve the predictive power, and, at least for the PAHs, the skeletal structure is a sufficient descriptor. 

\subsection{Application to specific PAHs}

Finally, we test the predictive power of the NNs in detail for three PAHs which have recently been discussed in terms of their presence in the interstellar medium \citep{McGuire2018,Bouwman2019}. Two of these, perylene and peropyrene \citep{Bouwman2019}, are present in the NASA spectroscopic database, while one, benzoniztrile \citep{McGuire2018}, is not. Benzonitrile is furthermore very different from the other molecules in the dataset from a structural point of view: first of all, it contains a \ce{C#N} triple bond which is not present in any molecule in the dataset and, secondly, the database is focused on PAHs so there is only another single aromatic ring in the dataset (phenol). To show how our method performs for those compounds, we trained $20$ additional NNs where perylene and peropyrene were explicitly included in the test set and thus excluded from the training material. Furthermore, we calculated the vibrational spectrum of benzonitrile with DFT using the same prescription employed for the compounds in the database and described in the methodological section.

\begin{figure}[ht]
    \centering
    \includegraphics[width = 0.45\textwidth]{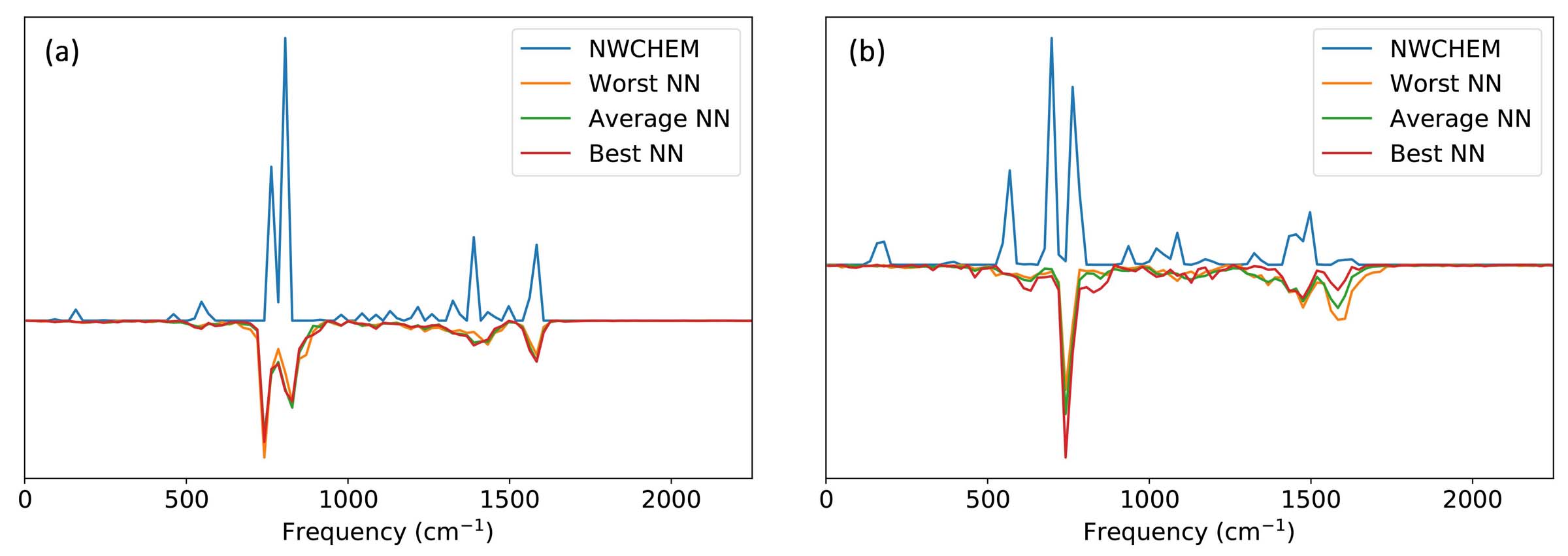}
    \caption{Worst, average and best neural network predictions for perylene and benzonitrile compared to the database spectra.}
    \label{fig:Perylene}
\end{figure}

The EMDs for perylene and peropyrene are $1.44$ and $2.29$ respectively, both in the same range as the EMD caused by using a different basis set for perylene, Fig.~\ref{fig:cross_method_emd}. The good quality of these predictions is illustrated in Fig.~\ref{fig:Perylene}a where it can be seen that the main peaks are found at the right frequencies. As expected, a much larger EMD of $9.56$ is found for benzonitrile. This would indicate a poor agreement between the calculated spectra and the model spectra, also illustrated in Fig.~\ref{fig:Perylene}b. As a regression model, the trained NN regression is inadequate for use on materials that differ substantially from the training set.

\subsection{Error estimation}

It is important that the model also be able to identify such cases. 
As a measure of uncertainty we have defined the cross-NN EMD as the average EMD between every prediction provided by an ensemble of NNs for a given molecule 

\begin{equation*}
    \frac{2}{N(N-1)}\sum_{i \ne j} \mathrm{EMD}\left(\mathrm{NN}_i,\mathrm{NN}_j\right)
\end{equation*}

Where $\mathrm{NN}_i$ is the spectrum predicted by the $i$-th neural network in the ensemble for the molecule and N is the number of NNs in the ensemble.

\begin{figure}[ht]
    \centering
  \includegraphics[width=.4\textwidth]{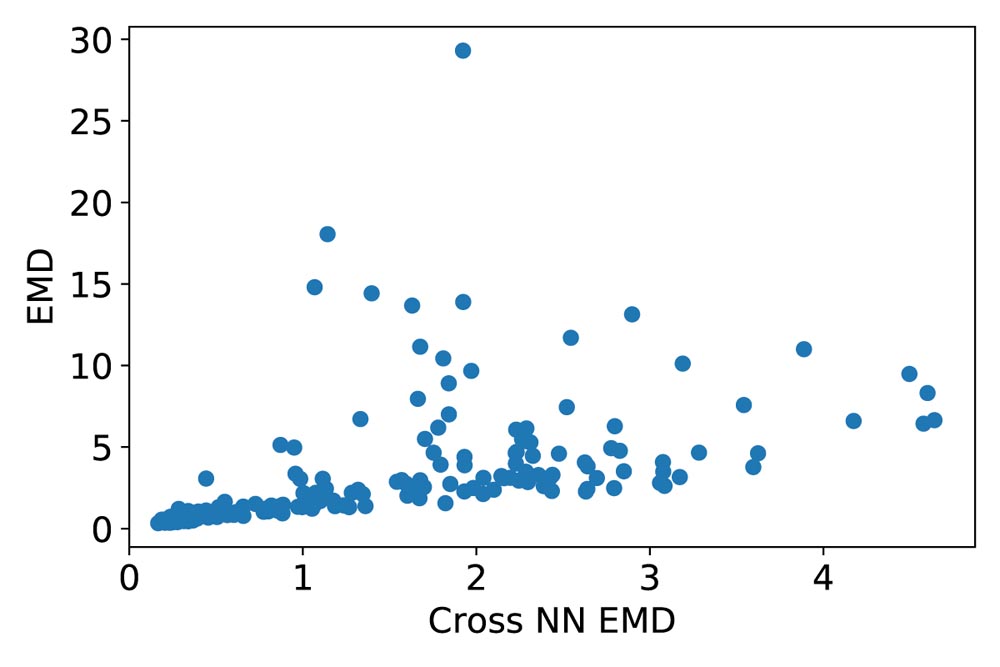}
  \caption{Average EMD from the NN predictions to the database spectrum vs the average cross EMD between the ensemble of NN predictions. }
  \label{fig:NN_cross_emd}
\end{figure}

The poor predictive power for benzonitrile is clearly reflected in these cross-validation EMD. The average and largest cross-NN EMD are $2.54$ and $9.30$ respectively. In contrast, for perylene and peropyrene the average/largest cross-NN EMDs are $0.66/1.53$ and $0.76/1.60$ respectively. In general, this tendency can also be shown for the test set of the original database. In Fig~\ref{fig:NN_cross_emd} a clear correlation between the average cross EMD and the error compared to the true value can be appreciated. Roughly speaking, we can identify three areas: an average cross EMD below $1$ points to a large probability that the predicted spectrum is reliable, an average cross EMD between $1$ and $2.5$ indicates that the model prediction might still be correct, and an average cross EMD larger than $2.5$ is correlated with a model prediction that is probably incorrect.

\section{Conclusions}

We extract the set of molecular structures from the NASA AMES PAH IR repository and translate them into topological fingerprints identifying the abundance of different chemical fragments in their molecular graph. We also extract the infrared spectra from the database, codify them into histograms and split them into a low- and a high-frequency part. Using the earth mover's distance as a metric we design and train a multilayer NN model to predict each of those parts of the spectrum based on the fingerprints. The resulting models show excellent predictive power for out-of-sample IR spectra, making them suitable for predicting the spectra of larger libraries of PAHs that better support more accurate interpretations of the astronomical IR observations. Moreover, NNs are able to recover identifiable pieces of knowledge, like the role of hydrogens in high-frequency vibrations. This will be helpful for answering the puzzling questions raised by astronomical observations on the chemical composition of ISM \citep{Li2020}.

We compare the NN predictions with DFT calculations for three different compounds and show that the average error of the NN predictions falls in the range of errors caused by choosing different basis set for the DFT calculation. We also find that the NN is only applicable to compounds similar to the training set and use multiple NNs to give an approximation of the expected error of the predictions.

We complement this analysis using RF regression. While the accuracy of RFs is lower than what can be achieved with NNs, they allow us to explore which molecular features are most relevant for determining the molecular spectra. We identify four substructures of high importance for both the low- and high-frequency parts of the spectrum. At the same time, however, the results point to a high degree of fungibility among descriptors, whereby similar levels of performance can be achieved using different combinations of those. We also check whether any important information about the molecule is left out by the topological descriptors by supplementing them with geometric information in the form of the largest eigenvalues of the Coulomb matrix. The models do not improve to any significant degree, showing that the topology of the molecular graph alone is enough to satisfactorily characterize the vibrational dynamics of these structures.

This study shows that NNs can be efficiently trained to bypass expensive first-principles calculations, offering useful levels of accuracy and incomparably lower computational cost even for demanding properties like the vibrational spectra. Moreover, it points to the possibility of extracting simple, intuitive rules from trained models that replicate or supplement existing specialist knowledge.\\

\textbf{Data access}: An example model, dataset and training code is available on Zenodo \citep{code}.


\begin{thebibliography}{61}
\expandafter\ifx\csname natexlab\endcsname\relax\def\natexlab#1{#1}\fi
\expandafter\ifx\csname bibnamefont\endcsname\relax
  \def\bibnamefont#1{#1}\fi
\expandafter\ifx\csname bibfnamefont\endcsname\relax
  \def\bibfnamefont#1{#1}\fi
\expandafter\ifx\csname citenamefont\endcsname\relax
  \def\citenamefont#1{#1}\fi
\expandafter\ifx\csname url\endcsname\relax
  \def\url#1{\texttt{#1}}\fi
\expandafter\ifx\csname urlprefix\endcsname\relax\def\urlprefix{URL }\fi
\providecommand{\bibinfo}[2]{#2}
\providecommand{\eprint}[2][]{\url{#2}}

\bibitem[{\citenamefont{{Herbst} and {van Dishoeck}}(2009)}]{Herbst2009}
\bibinfo{author}{\bibfnamefont{E.}~\bibnamefont{{Herbst}}} \bibnamefont{and}
  \bibinfo{author}{\bibfnamefont{E.~F.} \bibnamefont{{van Dishoeck}}},
  \bibinfo{journal}{ARA\&A} \textbf{\bibinfo{volume}{47}}, \bibinfo{pages}{427}
  (\bibinfo{year}{2009}).

\bibitem[{\citenamefont{{Zhang} et~al.}(2015)\citenamefont{{Zhang}, {Cao},
  {Colella}, {Liang}, {Bredas}, {Houk}, and {Briseno}}}]{Zhang2014}
\bibinfo{author}{\bibfnamefont{L.}~\bibnamefont{{Zhang}}},
  \bibinfo{author}{\bibfnamefont{Y.}~\bibnamefont{{Cao}}},
  \bibinfo{author}{\bibfnamefont{N.~S.} \bibnamefont{{Colella}}},
  \bibinfo{author}{\bibfnamefont{Y.}~\bibnamefont{{Liang}}},
  \bibinfo{author}{\bibfnamefont{J.}~\bibnamefont{{Bredas}}},
  \bibinfo{author}{\bibfnamefont{K.~N.} \bibnamefont{{Houk}}},
  \bibnamefont{and} \bibinfo{author}{\bibfnamefont{A.~L.}
  \bibnamefont{{Briseno}}}, \bibinfo{journal}{Acc. Chem. Res.}
  \textbf{\bibinfo{volume}{48}}, \bibinfo{pages}{500} (\bibinfo{year}{2015}).

\bibitem[{\citenamefont{{Moorthy} et~al.}(2015)\citenamefont{{Moorthy}, {Chu},
  and {Carlin}}}]{Moorthy2015}
\bibinfo{author}{\bibfnamefont{B.}~\bibnamefont{{Moorthy}}},
  \bibinfo{author}{\bibfnamefont{C.}~\bibnamefont{{Chu}}}, \bibnamefont{and}
  \bibinfo{author}{\bibfnamefont{D.~J.} \bibnamefont{{Carlin}}},
  \bibinfo{journal}{Toxicol. Sci.} \textbf{\bibinfo{volume}{145}},
  \bibinfo{pages}{5} (\bibinfo{year}{2015}).

\bibitem[{\citenamefont{{Ravindra} et~al.}(2008)\citenamefont{{Ravindra},
  {Sokhi}, and {van Grieken}}}]{Ravindra2008}
\bibinfo{author}{\bibfnamefont{K.}~\bibnamefont{{Ravindra}}},
  \bibinfo{author}{\bibfnamefont{R.}~\bibnamefont{{Sokhi}}}, \bibnamefont{and}
  \bibinfo{author}{\bibfnamefont{R.}~\bibnamefont{{van Grieken}}},
  \bibinfo{journal}{AtmEn} \textbf{\bibinfo{volume}{42}}, \bibinfo{pages}{2895}
  (\bibinfo{year}{2008}).

\bibitem[{\citenamefont{{Snow} and {Witt}}(1995)}]{Snow1995}
\bibinfo{author}{\bibfnamefont{T.~P.} \bibnamefont{{Snow}}} \bibnamefont{and}
  \bibinfo{author}{\bibfnamefont{A.~N.} \bibnamefont{{Witt}}},
  \bibinfo{journal}{Sci} \textbf{\bibinfo{volume}{270}}, \bibinfo{pages}{1455}
  (\bibinfo{year}{1995}).

\bibitem[{\citenamefont{{Tielens}}(2008)}]{Tielens2008}
\bibinfo{author}{\bibfnamefont{A.~G. G.~M.} \bibnamefont{{Tielens}}},
  \bibinfo{journal}{ARA\&A} \textbf{\bibinfo{volume}{46}}, \bibinfo{pages}{289}
  (\bibinfo{year}{2008}).

\bibitem[{\citenamefont{{Hardegree-Ullman}
  et~al.}(2014)\citenamefont{{Hardegree-Ullman}, {Gudipati}, {Boogert},
  {Lignell}, {Allamandola}, {Stapelfeldt}, and {Werner}}}]{Hardegree2014}
\bibinfo{author}{\bibfnamefont{E.~E.} \bibnamefont{{Hardegree-Ullman}}},
  \bibinfo{author}{\bibfnamefont{M.~S.} \bibnamefont{{Gudipati}}},
  \bibinfo{author}{\bibfnamefont{A.~C.~A.} \bibnamefont{{Boogert}}},
  \bibinfo{author}{\bibfnamefont{H.}~\bibnamefont{{Lignell}}},
  \bibinfo{author}{\bibfnamefont{L.~J.} \bibnamefont{{Allamandola}}},
  \bibinfo{author}{\bibfnamefont{K.~R.} \bibnamefont{{Stapelfeldt}}},
  \bibnamefont{and} \bibinfo{author}{\bibfnamefont{M.}~\bibnamefont{{Werner}}},
  \bibinfo{journal}{ApJ} \textbf{\bibinfo{volume}{784}}, \bibinfo{pages}{172}
  (\bibinfo{year}{2014}).

\bibitem[{\citenamefont{{McGuire} et~al.}(2018)\citenamefont{{McGuire},
  {Burkhardt}, {Kalenskii}, {Shingledecker}, {Remijan}, {Herbst}, and
  {McCarthy}}}]{McGuire2018}
\bibinfo{author}{\bibfnamefont{B.~A.} \bibnamefont{{McGuire}}},
  \bibinfo{author}{\bibfnamefont{A.~M.} \bibnamefont{{Burkhardt}}},
  \bibinfo{author}{\bibfnamefont{S.}~\bibnamefont{{Kalenskii}}},
  \bibinfo{author}{\bibfnamefont{C.~N.} \bibnamefont{{Shingledecker}}},
  \bibinfo{author}{\bibfnamefont{A.~J.} \bibnamefont{{Remijan}}},
  \bibinfo{author}{\bibfnamefont{E.}~\bibnamefont{{Herbst}}}, \bibnamefont{and}
  \bibinfo{author}{\bibfnamefont{M.~C.} \bibnamefont{{McCarthy}}},
  \bibinfo{journal}{Sci} \textbf{\bibinfo{volume}{359}}, \bibinfo{pages}{202}
  (\bibinfo{year}{2018}).

\bibitem[{\citenamefont{{Qi} et~al.}(2018)\citenamefont{{Qi}, {Picaud},
  {Devel}, {Liang}, and {Wang}}}]{Qihn2018}
\bibinfo{author}{\bibfnamefont{H.}~\bibnamefont{{Qi}}},
  \bibinfo{author}{\bibfnamefont{S.}~\bibnamefont{{Picaud}}},
  \bibinfo{author}{\bibfnamefont{M.}~\bibnamefont{{Devel}}},
  \bibinfo{author}{\bibfnamefont{E.}~\bibnamefont{{Liang}}}, \bibnamefont{and}
  \bibinfo{author}{\bibfnamefont{Z.}~\bibnamefont{{Wang}}},
  \bibinfo{journal}{ApJ} \textbf{\bibinfo{volume}{867}}, \bibinfo{pages}{133}
  (\bibinfo{year}{2018}).

\bibitem[{\citenamefont{{Ehrenfreund} and {Charnley}}(2000)}]{Ehrenfreund2000}
\bibinfo{author}{\bibfnamefont{P.}~\bibnamefont{{Ehrenfreund}}}
  \bibnamefont{and} \bibinfo{author}{\bibfnamefont{S.~B.}
  \bibnamefont{{Charnley}}}, \bibinfo{journal}{ARA\&A}
  \textbf{\bibinfo{volume}{38}}, \bibinfo{pages}{427} (\bibinfo{year}{2000}).

\bibitem[{\citenamefont{{Neubrech} et~al.}(2017)\citenamefont{{Neubrech},
  {Huck}, {Weber}, {Pucci}, and {Giessen}}}]{Neubrech2017}
\bibinfo{author}{\bibfnamefont{F.}~\bibnamefont{{Neubrech}}},
  \bibinfo{author}{\bibfnamefont{C.}~\bibnamefont{{Huck}}},
  \bibinfo{author}{\bibfnamefont{K.}~\bibnamefont{{Weber}}},
  \bibinfo{author}{\bibfnamefont{A.}~\bibnamefont{{Pucci}}}, \bibnamefont{and}
  \bibinfo{author}{\bibfnamefont{H.}~\bibnamefont{{Giessen}}},
  \bibinfo{journal}{Chem. Rev.} \textbf{\bibinfo{volume}{117}},
  \bibinfo{pages}{5110} (\bibinfo{year}{2017}).

\bibitem[{\citenamefont{{Meier}}(2003)}]{Meier2003}
\bibinfo{author}{\bibfnamefont{R.}~\bibnamefont{{Meier}}},
  \emph{\bibinfo{title}{{Handbook of vibrational spectroscopy}}}
  (\bibinfo{publisher}{John Wiley \& Sons Ltd. : Chichester, UK},
  \bibinfo{year}{2003}).

\bibitem[{\citenamefont{{Young} et~al.}(2012)\citenamefont{{Young}, {Becklin},
  {Marcum}, {Roellig}, {Buizer}, and {Herter}}}]{Young2012}
\bibinfo{author}{\bibfnamefont{E.~T.} \bibnamefont{{Young}}},
  \bibinfo{author}{\bibfnamefont{E.~E.} \bibnamefont{{Becklin}}},
  \bibinfo{author}{\bibfnamefont{P.~M.} \bibnamefont{{Marcum}}},
  \bibinfo{author}{\bibfnamefont{T.~L.} \bibnamefont{{Roellig}}},
  \bibinfo{author}{\bibfnamefont{J.~M.~D.} \bibnamefont{{Buizer}}},
  \bibnamefont{and} \bibinfo{author}{\bibfnamefont{T.~L.}
  \bibnamefont{{Herter}}}, \bibinfo{journal}{ApJL}
  \textbf{\bibinfo{volume}{749}}, \bibinfo{pages}{L17} (\bibinfo{year}{2012}).

\bibitem[{\citenamefont{{Deming} and {Knutson}}(2020)}]{Deming2020}
\bibinfo{author}{\bibfnamefont{D.}~\bibnamefont{{Deming}}} \bibnamefont{and}
  \bibinfo{author}{\bibfnamefont{H.~A.} \bibnamefont{{Knutson}}},
  \bibinfo{journal}{NatAs} \textbf{\bibinfo{volume}{4}}, \bibinfo{pages}{453}
  (\bibinfo{year}{2020}).

\bibitem[{\citenamefont{{Croiset} et~al.}(2016)\citenamefont{{Croiset},
  {Candian}, {Berne}, and {Tielens}}}]{Croiset2016}
\bibinfo{author}{\bibfnamefont{B.~A.} \bibnamefont{{Croiset}}},
  \bibinfo{author}{\bibfnamefont{A.}~\bibnamefont{{Candian}}},
  \bibinfo{author}{\bibfnamefont{O.}~\bibnamefont{{Berne}}}, \bibnamefont{and}
  \bibinfo{author}{\bibfnamefont{A.~G. G.~M.} \bibnamefont{{Tielens}}},
  \bibinfo{journal}{A\&A} \textbf{\bibinfo{volume}{590}}, \bibinfo{pages}{A26}
  (\bibinfo{year}{2016}).

\bibitem[{\citenamefont{{Andrews} et~al.}(2015)\citenamefont{{Andrews},
  {Boersma}, {Werner}, {Livingston}, {Allamandola}, and
  {Tielens}}}]{Andrews2015}
\bibinfo{author}{\bibfnamefont{H.}~\bibnamefont{{Andrews}}},
  \bibinfo{author}{\bibfnamefont{C.}~\bibnamefont{{Boersma}}},
  \bibinfo{author}{\bibfnamefont{M.~W.} \bibnamefont{{Werner}}},
  \bibinfo{author}{\bibfnamefont{J.}~\bibnamefont{{Livingston}}},
  \bibinfo{author}{\bibfnamefont{L.~J.} \bibnamefont{{Allamandola}}},
  \bibnamefont{and} \bibinfo{author}{\bibfnamefont{A.~G. G.~M.}
  \bibnamefont{{Tielens}}}, \bibinfo{journal}{ApJ}
  \textbf{\bibinfo{volume}{807}}, \bibinfo{pages}{99} (\bibinfo{year}{2015}).

\bibitem[{\citenamefont{{Shannon} et~al.}(2018)\citenamefont{{Shannon},
  {Peeters}, {Cami}, and {Blommaert}}}]{Shannon2018}
\bibinfo{author}{\bibfnamefont{M.~J.} \bibnamefont{{Shannon}}},
  \bibinfo{author}{\bibfnamefont{E.}~\bibnamefont{{Peeters}}},
  \bibinfo{author}{\bibfnamefont{J.}~\bibnamefont{{Cami}}}, \bibnamefont{and}
  \bibinfo{author}{\bibfnamefont{J.~A. D.~L.} \bibnamefont{{Blommaert}}},
  \bibinfo{journal}{ApJ} \textbf{\bibinfo{volume}{855}}, \bibinfo{pages}{32}
  (\bibinfo{year}{2018}).

\bibitem[{\citenamefont{{Allamandola} et~al.}(1999)\citenamefont{{Allamandola},
  {Hudgins}, and {Sandford}}}]{Allamandola1999}
\bibinfo{author}{\bibfnamefont{L.~J.} \bibnamefont{{Allamandola}}},
  \bibinfo{author}{\bibfnamefont{D.~M.} \bibnamefont{{Hudgins}}},
  \bibnamefont{and} \bibinfo{author}{\bibfnamefont{S.~A.}
  \bibnamefont{{Sandford}}}, \bibinfo{journal}{ApJ}
  \textbf{\bibinfo{volume}{511}}, \bibinfo{pages}{L115} (\bibinfo{year}{1999}).

\bibitem[{\citenamefont{{Maltseva} et~al.}(2015)\citenamefont{{Maltseva},
  {Petrignani}, {Candian}, {Mackie}, {Huang}, {Lee}, {Tielens}, {Oomens}, and
  {Buma}}}]{Maltseva2015}
\bibinfo{author}{\bibfnamefont{E.}~\bibnamefont{{Maltseva}}},
  \bibinfo{author}{\bibfnamefont{A.}~\bibnamefont{{Petrignani}}},
  \bibinfo{author}{\bibfnamefont{A.}~\bibnamefont{{Candian}}},
  \bibinfo{author}{\bibfnamefont{C.~J.} \bibnamefont{{Mackie}}},
  \bibinfo{author}{\bibfnamefont{X.~C.} \bibnamefont{{Huang}}},
  \bibinfo{author}{\bibfnamefont{T.~J.} \bibnamefont{{Lee}}},
  \bibinfo{author}{\bibfnamefont{A.~G. G.~M.} \bibnamefont{{Tielens}}},
  \bibinfo{author}{\bibfnamefont{J.}~\bibnamefont{{Oomens}}}, \bibnamefont{and}
  \bibinfo{author}{\bibfnamefont{W.~J.} \bibnamefont{{Buma}}},
  \bibinfo{journal}{ApJ} \textbf{\bibinfo{volume}{814}}, \bibinfo{pages}{23}
  (\bibinfo{year}{2015}).

\bibitem[{\citenamefont{{Bouwman} et~al.}(2019)\citenamefont{{Bouwman},
  {Castellanos}, {Bulak}, {Scheltinga}, {Cami}, and {Tielens}}}]{Bouwman2019}
\bibinfo{author}{\bibfnamefont{J.}~\bibnamefont{{Bouwman}}},
  \bibinfo{author}{\bibfnamefont{P.}~\bibnamefont{{Castellanos}}},
  \bibinfo{author}{\bibfnamefont{M.}~\bibnamefont{{Bulak}}},
  \bibinfo{author}{\bibfnamefont{J.~T.~V.} \bibnamefont{{Scheltinga}}},
  \bibinfo{author}{\bibfnamefont{H.}~\bibnamefont{{Cami}},
  \bibfnamefont{J.~Linnartz}}, \bibnamefont{and}
  \bibinfo{author}{\bibfnamefont{A.~G. G.~M.} \bibnamefont{{Tielens}}},
  \bibinfo{journal}{A\&A} \textbf{\bibinfo{volume}{321}}, \bibinfo{pages}{A80}
  (\bibinfo{year}{2019}).

\bibitem[{\citenamefont{{Kwok} and {Zhang}}(2011)}]{Kwok2011}
\bibinfo{author}{\bibfnamefont{S.}~\bibnamefont{{Kwok}}} \bibnamefont{and}
  \bibinfo{author}{\bibfnamefont{Y.}~\bibnamefont{{Zhang}}},
  \bibinfo{journal}{Natur} \textbf{\bibinfo{volume}{479}}, \bibinfo{pages}{80}
  (\bibinfo{year}{2011}).

\bibitem[{\citenamefont{{Li} and {Draine}}(2012)}]{Li2012}
\bibinfo{author}{\bibfnamefont{A.}~\bibnamefont{{Li}}} \bibnamefont{and}
  \bibinfo{author}{\bibfnamefont{B.~T.} \bibnamefont{{Draine}}},
  \bibinfo{journal}{ApJL} \textbf{\bibinfo{volume}{760}}, \bibinfo{pages}{L35}
  (\bibinfo{year}{2012}).

\bibitem[{\citenamefont{{Kwok} and {Zhang}}(2013)}]{Kwok2013}
\bibinfo{author}{\bibfnamefont{S.}~\bibnamefont{{Kwok}}} \bibnamefont{and}
  \bibinfo{author}{\bibfnamefont{Y.}~\bibnamefont{{Zhang}}},
  \bibinfo{journal}{ApJ} \textbf{\bibinfo{volume}{771}}, \bibinfo{pages}{5}
  (\bibinfo{year}{2013}).

\bibitem[{\citenamefont{{Butler} et~al.}(2018)\citenamefont{{Butler}, {Davies},
  {Cartwright}, {Isayev}, and {Walsh}}}]{Butler2018}
\bibinfo{author}{\bibfnamefont{K.~T.} \bibnamefont{{Butler}}},
  \bibinfo{author}{\bibfnamefont{D.~W.} \bibnamefont{{Davies}}},
  \bibinfo{author}{\bibfnamefont{H.}~\bibnamefont{{Cartwright}}},
  \bibinfo{author}{\bibfnamefont{O.}~\bibnamefont{{Isayev}}}, \bibnamefont{and}
  \bibinfo{author}{\bibfnamefont{A.}~\bibnamefont{{Walsh}}},
  \bibinfo{journal}{Natur} \textbf{\bibinfo{volume}{559}}, \bibinfo{pages}{547}
  (\bibinfo{year}{2018}).

\bibitem[{\citenamefont{{Marquez-Neila}
  et~al.}(2018)\citenamefont{{Marquez-Neila}, {Fisher}, {Sznitman}, and
  {Heng}}}]{Marquez2018}
\bibinfo{author}{\bibfnamefont{P.}~\bibnamefont{{Marquez-Neila}}},
  \bibinfo{author}{\bibfnamefont{C.}~\bibnamefont{{Fisher}}},
  \bibinfo{author}{\bibfnamefont{R.}~\bibnamefont{{Sznitman}}},
  \bibnamefont{and} \bibinfo{author}{\bibfnamefont{K.}~\bibnamefont{{Heng}}},
  \bibinfo{journal}{NatAs} \textbf{\bibinfo{volume}{2}}, \bibinfo{pages}{719}
  (\bibinfo{year}{2018}).

\bibitem[{\citenamefont{{Ghosh} et~al.}(2019)\citenamefont{{Ghosh}, {Stuke},
  {Todorovi\v{c}}, {J{\o}rgensen}, {Schmidt}, {Vehtari}, and
  {Rinke}}}]{Ghosh2019}
\bibinfo{author}{\bibfnamefont{K.}~\bibnamefont{{Ghosh}}},
  \bibinfo{author}{\bibfnamefont{A.}~\bibnamefont{{Stuke}}},
  \bibinfo{author}{\bibfnamefont{M.}~\bibnamefont{{Todorovi\v{c}}}},
  \bibinfo{author}{\bibfnamefont{P.~B.} \bibnamefont{{J{\o}rgensen}}},
  \bibinfo{author}{\bibfnamefont{M.~N.} \bibnamefont{{Schmidt}}},
  \bibinfo{author}{\bibfnamefont{A.}~\bibnamefont{{Vehtari}}},
  \bibnamefont{and} \bibinfo{author}{\bibfnamefont{P.}~\bibnamefont{{Rinke}}},
  \bibinfo{journal}{Adv. Sci.} \textbf{\bibinfo{volume}{6}},
  \bibinfo{pages}{1801367} (\bibinfo{year}{2019}).

\bibitem[{\citenamefont{{Gastegger} et~al.}(2017)\citenamefont{{Gastegger},
  {Behler}, and {Marquetand}}}]{Gastegger2017}
\bibinfo{author}{\bibfnamefont{M.}~\bibnamefont{{Gastegger}}},
  \bibinfo{author}{\bibfnamefont{J.}~\bibnamefont{{Behler}}}, \bibnamefont{and}
  \bibinfo{author}{\bibfnamefont{P.}~\bibnamefont{{Marquetand}}},
  \bibinfo{journal}{Chem. Sci.} \textbf{\bibinfo{volume}{8}},
  \bibinfo{pages}{6924} (\bibinfo{year}{2017}).

\bibitem[{\citenamefont{{Weigel} and {Herges}}(1996)}]{Weigel1996}
\bibinfo{author}{\bibfnamefont{U.~M.} \bibnamefont{{Weigel}}} \bibnamefont{and}
  \bibinfo{author}{\bibfnamefont{R.}~\bibnamefont{{Herges}}},
  \bibinfo{journal}{Anal. Chim. Acta} \textbf{\bibinfo{volume}{331}},
  \bibinfo{pages}{63} (\bibinfo{year}{1996}).

\bibitem[{\citenamefont{{Selzer} et~al.}(2000)\citenamefont{{Selzer},
  {Gasteiger}, {Thomas}, and {Salzer}}}]{Selzer2000}
\bibinfo{author}{\bibfnamefont{P.}~\bibnamefont{{Selzer}}},
  \bibinfo{author}{\bibfnamefont{J.}~\bibnamefont{{Gasteiger}}},
  \bibinfo{author}{\bibfnamefont{H.}~\bibnamefont{{Thomas}}}, \bibnamefont{and}
  \bibinfo{author}{\bibfnamefont{R.}~\bibnamefont{{Salzer}}},
  \bibinfo{journal}{Chem. Eur. J.} \textbf{\bibinfo{volume}{6}},
  \bibinfo{pages}{920} (\bibinfo{year}{2000}).

\bibitem[{\citenamefont{{Rupp} et~al.}(2012)\citenamefont{{Rupp}, {Tkatchenko},
  {M\"uller}, and {von Lilienfeld}}}]{Rupp2012}
\bibinfo{author}{\bibfnamefont{M.}~\bibnamefont{{Rupp}}},
  \bibinfo{author}{\bibfnamefont{A.}~\bibnamefont{{Tkatchenko}}},
  \bibinfo{author}{\bibfnamefont{K.~R.} \bibnamefont{{M\"uller}}},
  \bibnamefont{and} \bibinfo{author}{\bibfnamefont{O.~A.} \bibnamefont{{von
  Lilienfeld}}}, \bibinfo{journal}{PhRvL} \textbf{\bibinfo{volume}{108}},
  \bibinfo{pages}{058301} (\bibinfo{year}{2012}).

\bibitem[{\citenamefont{{Sch\"utt} et~al.}(2014)\citenamefont{{Sch\"utt},
  {Glawe}, {Brockherde}, {Sanna}, {M\"uller}, and {Gross}}}]{Schutt2014}
\bibinfo{author}{\bibfnamefont{K.~T.} \bibnamefont{{Sch\"utt}}},
  \bibinfo{author}{\bibfnamefont{H.}~\bibnamefont{{Glawe}}},
  \bibinfo{author}{\bibfnamefont{F.}~\bibnamefont{{Brockherde}}},
  \bibinfo{author}{\bibfnamefont{A.}~\bibnamefont{{Sanna}}},
  \bibinfo{author}{\bibfnamefont{K.~R.} \bibnamefont{{M\"uller}}},
  \bibnamefont{and} \bibinfo{author}{\bibfnamefont{E.~K.~U.}
  \bibnamefont{{Gross}}}, \bibinfo{journal}{PhRvB}
  \textbf{\bibinfo{volume}{89}}, \bibinfo{pages}{205118}
  (\bibinfo{year}{2014}).

\bibitem[{\citenamefont{{Bauschlicher}
  et~al.}(2010)\citenamefont{{Bauschlicher}, {Boersma}, {Ricca}, {Mattioda},
  {Cami}, {Peeters}, {de Armas}, {Saborido}, {Hudgins}, and
  {Allamandola}}}]{Bauschlicher2010}
\bibinfo{author}{\bibfnamefont{C.~W.~J.} \bibnamefont{{Bauschlicher}}},
  \bibinfo{author}{\bibfnamefont{C.}~\bibnamefont{{Boersma}}},
  \bibinfo{author}{\bibfnamefont{A.}~\bibnamefont{{Ricca}}},
  \bibinfo{author}{\bibfnamefont{A.~L.} \bibnamefont{{Mattioda}}},
  \bibinfo{author}{\bibfnamefont{J.}~\bibnamefont{{Cami}}},
  \bibinfo{author}{\bibfnamefont{E.}~\bibnamefont{{Peeters}}},
  \bibinfo{author}{\bibfnamefont{F.~S.} \bibnamefont{{de Armas}}},
  \bibinfo{author}{\bibfnamefont{G.~P.} \bibnamefont{{Saborido}}},
  \bibinfo{author}{\bibfnamefont{D.~M.} \bibnamefont{{Hudgins}}},
  \bibnamefont{and} \bibinfo{author}{\bibfnamefont{L.~J.}
  \bibnamefont{{Allamandola}}}, \bibinfo{journal}{ApJS}
  \textbf{\bibinfo{volume}{189}}, \bibinfo{pages}{341} (\bibinfo{year}{2010}).

\bibitem[{\citenamefont{{Boersma} et~al.}(2014)\citenamefont{{Boersma},
  {Bauschlicher}, {Ricca}, {Mattioda}, {Cami}, {Peeters}, {Armas}, {Saborido},
  {Hudgins}, and {Allamandola}}}]{Boersma2014}
\bibinfo{author}{\bibfnamefont{C.}~\bibnamefont{{Boersma}}},
  \bibinfo{author}{\bibfnamefont{C.~W.~J.} \bibnamefont{{Bauschlicher}}},
  \bibinfo{author}{\bibfnamefont{A.}~\bibnamefont{{Ricca}}},
  \bibinfo{author}{\bibfnamefont{A.~L.} \bibnamefont{{Mattioda}}},
  \bibinfo{author}{\bibfnamefont{J.}~\bibnamefont{{Cami}}},
  \bibinfo{author}{\bibfnamefont{E.}~\bibnamefont{{Peeters}}},
  \bibinfo{author}{\bibfnamefont{F.~S.~d.} \bibnamefont{{Armas}}},
  \bibinfo{author}{\bibfnamefont{G.~P.} \bibnamefont{{Saborido}}},
  \bibinfo{author}{\bibfnamefont{D.~M.} \bibnamefont{{Hudgins}}},
  \bibnamefont{and} \bibinfo{author}{\bibfnamefont{L.~J.}
  \bibnamefont{{Allamandola}}}, \bibinfo{journal}{ApJS}
  \textbf{\bibinfo{volume}{211}}, \bibinfo{pages}{8} (\bibinfo{year}{2014}).

\bibitem[{\citenamefont{{Bauschlicher}
  et~al.}(2018)\citenamefont{{Bauschlicher}, {Ricca}, {Boersma}, and
  {Allamandola}}}]{Bauschlicher2018}
\bibinfo{author}{\bibfnamefont{C.~W.~J.} \bibnamefont{{Bauschlicher}}},
  \bibinfo{author}{\bibfnamefont{A.}~\bibnamefont{{Ricca}}},
  \bibinfo{author}{\bibfnamefont{C.}~\bibnamefont{{Boersma}}},
  \bibnamefont{and} \bibinfo{author}{\bibfnamefont{L.~J.}
  \bibnamefont{{Allamandola}}}, \bibinfo{journal}{ApJS}
  \textbf{\bibinfo{volume}{234}}, \bibinfo{pages}{32} (\bibinfo{year}{2018}).

\bibitem[{\citenamefont{{Allamandola} et~al.}(1989)\citenamefont{{Allamandola},
  {Tielens}, and {Barker}}}]{Allamandola1989}
\bibinfo{author}{\bibfnamefont{L.~J.} \bibnamefont{{Allamandola}}},
  \bibinfo{author}{\bibfnamefont{A.~G. G.~M.} \bibnamefont{{Tielens}}},
  \bibnamefont{and} \bibinfo{author}{\bibfnamefont{J.~R.}
  \bibnamefont{{Barker}}}, \bibinfo{journal}{ApJS}
  \textbf{\bibinfo{volume}{71}}, \bibinfo{pages}{733} (\bibinfo{year}{1989}).

\bibitem[{\citenamefont{{Draine} and {Li}}(2007)}]{Draine2007}
\bibinfo{author}{\bibfnamefont{B.~T.} \bibnamefont{{Draine}}} \bibnamefont{and}
  \bibinfo{author}{\bibfnamefont{A.}~\bibnamefont{{Li}}},
  \bibinfo{journal}{ApJ} \textbf{\bibinfo{volume}{657}}, \bibinfo{pages}{810}
  (\bibinfo{year}{2007}).

\bibitem[{\citenamefont{{Peeters}}(2011)}]{Peeters2011}
\bibinfo{author}{\bibfnamefont{E.}~\bibnamefont{{Peeters}}},
  \bibinfo{journal}{ProcIntAstroUni} \textbf{\bibinfo{volume}{280}},
  \bibinfo{pages}{149} (\bibinfo{year}{2011}).

\bibitem[{\citenamefont{Knuth}(2006)}]{Knuth2006}
\bibinfo{author}{\bibfnamefont{K.~H.} \bibnamefont{Knuth}},
  \emph{\bibinfo{title}{Optimal data-based binning for histograms}}
  (\bibinfo{year}{2006}), \eprint{arXiv:physics/0605197}.

\bibitem[{\citenamefont{{Monge}}(1781)}]{Monge1781}
\bibinfo{author}{\bibfnamefont{G.}~\bibnamefont{{Monge}}},
  \bibinfo{journal}{HARSB} pp. \bibinfo{pages}{666--704}
  (\bibinfo{year}{1781}),
  \eprint{https://gallica.bnf.fr/ark:/12148/bpt6k35800/f796}.

\bibitem[{\citenamefont{{Dobrushin}}(1970)}]{Dobrushin1970}
\bibinfo{author}{\bibfnamefont{R.~L.} \bibnamefont{{Dobrushin}}},
  \bibinfo{journal}{Theory Probab. its Appl.} p. \bibinfo{pages}{458}
  (\bibinfo{year}{1970}).

\bibitem[{\citenamefont{{Smirnov}}(1944)}]{Smirnov1944}
\bibinfo{author}{\bibfnamefont{N.~V.} \bibnamefont{{Smirnov}}},
  \bibinfo{journal}{Uspekhi Mat. Nauk} \textbf{\bibinfo{volume}{10}},
  \bibinfo{pages}{179} (\bibinfo{year}{1944}),
  \eprint{http://mi.mathnet.ru/eng/umn/y1944/i10/p179}.

\bibitem[{\citenamefont{{Valiev} et~al.}(2010)\citenamefont{{Valiev},
  {Bylaska}, {Govind}, {Kowalski}, and {Straatsma}}}]{nwchem}
\bibinfo{author}{\bibfnamefont{M.}~\bibnamefont{{Valiev}}},
  \bibinfo{author}{\bibfnamefont{E.~J.} \bibnamefont{{Bylaska}}},
  \bibinfo{author}{\bibfnamefont{N.}~\bibnamefont{{Govind}}},
  \bibinfo{author}{\bibfnamefont{K.}~\bibnamefont{{Kowalski}}},
  \bibnamefont{and}
  \bibinfo{author}{\bibfnamefont{T.}~\bibnamefont{{Straatsma}}},
  \bibinfo{journal}{Comput. Phys. Commun.} \textbf{\bibinfo{volume}{181}},
  \bibinfo{pages}{1477} (\bibinfo{year}{2010}).

\bibitem[{\citenamefont{{Rogers} and {Mathew}}(2010)}]{Rogers2010}
\bibinfo{author}{\bibfnamefont{D.}~\bibnamefont{{Rogers}}} \bibnamefont{and}
  \bibinfo{author}{\bibfnamefont{H.}~\bibnamefont{{Mathew}}},
  \bibinfo{journal}{J. Chem. Inf. Model.} \textbf{\bibinfo{volume}{50}},
  \bibinfo{pages}{742} (\bibinfo{year}{2010}).

\bibitem[{\citenamefont{RDKit}()}]{rdkit}
\bibinfo{author}{\bibnamefont{RDKit}}, \bibinfo{note}{http://www.rdkit.org}.

\bibitem[{\citenamefont{{Morgan}}(1965)}]{Morgan1965}
\bibinfo{author}{\bibfnamefont{H.~L.} \bibnamefont{{Morgan}}},
  \bibinfo{journal}{J. Chem. Doc.} \textbf{\bibinfo{volume}{5}},
  \bibinfo{pages}{107} (\bibinfo{year}{1965}).

\bibitem[{\citenamefont{{Weininger} et~al.}(1989)\citenamefont{{Weininger},
  {Weininger}, and {Weininger}}}]{Weininger1989}
\bibinfo{author}{\bibfnamefont{D.}~\bibnamefont{{Weininger}}},
  \bibinfo{author}{\bibfnamefont{A.}~\bibnamefont{{Weininger}}},
  \bibnamefont{and} \bibinfo{author}{\bibfnamefont{J.~L.}
  \bibnamefont{{Weininger}}}, \bibinfo{journal}{J. Chem. Inf. Comput. Sci.}
  \textbf{\bibinfo{volume}{29}}, \bibinfo{pages}{97} (\bibinfo{year}{1989}).

\bibitem[{\citenamefont{{Dalby et al.}}(1992)}]{sdf}
\bibinfo{author}{\bibfnamefont{A.}~\bibnamefont{{Dalby et al.}}},
  \bibinfo{journal}{J. Chem. Inf. Comput. Sci.} \textbf{\bibinfo{volume}{32}},
  \bibinfo{pages}{244} (\bibinfo{year}{1992}).

\bibitem[{\citenamefont{{O'Boyle} et~al.}(2011)\citenamefont{{O'Boyle},
  {Banck}, {James}, {Morley}, {Vandermeersch}, and {Hutchison}}}]{Boyle2011}
\bibinfo{author}{\bibfnamefont{N.~M.} \bibnamefont{{O'Boyle}}},
  \bibinfo{author}{\bibfnamefont{M.}~\bibnamefont{{Banck}}},
  \bibinfo{author}{\bibfnamefont{C.~A.} \bibnamefont{{James}}},
  \bibinfo{author}{\bibfnamefont{C.}~\bibnamefont{{Morley}}},
  \bibinfo{author}{\bibfnamefont{T.}~\bibnamefont{{Vandermeersch}}},
  \bibnamefont{and} \bibinfo{author}{\bibfnamefont{G.~R.}
  \bibnamefont{{Hutchison}}}, \bibinfo{journal}{J. Cheminformatics}
  \textbf{\bibinfo{volume}{3}}, \bibinfo{pages}{33} (\bibinfo{year}{2011}).

\bibitem[{\citenamefont{{Bishop}}(1996)}]{Bishop1996}
\bibinfo{author}{\bibfnamefont{C.~M.} \bibnamefont{{Bishop}}},
  \emph{\bibinfo{title}{{Neural Networks for Pattern Recognition}}}
  (\bibinfo{publisher}{Oxford University Press, Inc.}, \bibinfo{address}{USA},
  \bibinfo{year}{1996}), ISBN \bibinfo{isbn}{0198538499}.

\bibitem[{\citenamefont{{Lecun} et~al.}(2015)\citenamefont{{Lecun}, {Bengio},
  and {Hinton}}}]{Lecun2015}
\bibinfo{author}{\bibfnamefont{Y.}~\bibnamefont{{Lecun}}},
  \bibinfo{author}{\bibfnamefont{Y.}~\bibnamefont{{Bengio}}}, \bibnamefont{and}
  \bibinfo{author}{\bibfnamefont{G.}~\bibnamefont{{Hinton}}},
  \bibinfo{journal}{Natur} \textbf{\bibinfo{volume}{521}}, \bibinfo{pages}{436}
  (\bibinfo{year}{2015}).

\bibitem[{\citenamefont{{Glorot} and {Bengio}}(2010)}]{Glorot2010}
\bibinfo{author}{\bibfnamefont{X.}~\bibnamefont{{Glorot}}} \bibnamefont{and}
  \bibinfo{author}{\bibfnamefont{Y.}~\bibnamefont{{Bengio}}}, in
  \emph{\bibinfo{booktitle}{Proceedings of the Thirteenth International
  Conference on Artificial Intelligence and Statistics}}, edited by
  \bibinfo{editor}{\bibfnamefont{Y.~W.} \bibnamefont{Teh}} \bibnamefont{and}
  \bibinfo{editor}{\bibfnamefont{M.}~\bibnamefont{Titterington}}
  (\bibinfo{publisher}{PMLR}, \bibinfo{address}{Chia Laguna Resort, Sardinia,
  Italy}, \bibinfo{year}{2010}), vol.~\bibinfo{volume}{9} of
  \emph{\bibinfo{series}{Proceedings of Machine Learning Research}}, pp.
  \bibinfo{pages}{249--256},
  \eprint{http://proceedings.mlr.press/v9/glorot10a/glorot10a.pdf}.

\bibitem[{\citenamefont{Kingma and Ba}(2015)}]{ADAM}
\bibinfo{author}{\bibfnamefont{D.~P.} \bibnamefont{Kingma}} \bibnamefont{and}
  \bibinfo{author}{\bibfnamefont{J.}~\bibnamefont{Ba}}, in
  \emph{\bibinfo{booktitle}{3rd International Conference on Learning
  Representations, {ICLR} 2015, San Diego, CA, USA, May 7-9, 2015, Conference
  Track Proceedings}} (\bibinfo{year}{2015}), \eprint{arXiv:1412.6980}.

\bibitem[{\citenamefont{{Abadi et al.}}(2015)}]{tensorflow2015-whitepaper}
\bibinfo{author}{\bibfnamefont{M.}~\bibnamefont{{Abadi et al.}}},
  \emph{\bibinfo{title}{{TensorFlow: Large-scale machine learning on
  heterogeneous systems}}} (\bibinfo{year}{2015}), \eprint{arXiv:1603.04467v2}.

\bibitem[{\citenamefont{{Breiman}}(2001)}]{random_forests}
\bibinfo{author}{\bibfnamefont{L.}~\bibnamefont{{Breiman}}},
  \bibinfo{journal}{Mach. Learn.} \textbf{\bibinfo{volume}{45}},
  \bibinfo{pages}{5} (\bibinfo{year}{2001}).

\bibitem[{\citenamefont{{Carrete} et~al.}(2014)\citenamefont{{Carrete}, {Li},
  {Mingo}, {Wang}, and {Curtarolo}}}]{PhysRevX.4.011019}
\bibinfo{author}{\bibfnamefont{J.}~\bibnamefont{{Carrete}}},
  \bibinfo{author}{\bibfnamefont{W.}~\bibnamefont{{Li}}},
  \bibinfo{author}{\bibfnamefont{N.}~\bibnamefont{{Mingo}}},
  \bibinfo{author}{\bibfnamefont{S.}~\bibnamefont{{Wang}}}, \bibnamefont{and}
  \bibinfo{author}{\bibfnamefont{S.}~\bibnamefont{{Curtarolo}}},
  \bibinfo{journal}{PhRvX} \textbf{\bibinfo{volume}{4}},
  \bibinfo{pages}{011019} (\bibinfo{year}{2014}).

\bibitem[{\citenamefont{{Pedregosa} et~al.}(2011)\citenamefont{{Pedregosa},
  {Varoquaux}, {Gramfort}, and {Michel}}}]{scikit-learn}
\bibinfo{author}{\bibfnamefont{F.}~\bibnamefont{{Pedregosa}}},
  \bibinfo{author}{\bibfnamefont{G.}~\bibnamefont{{Varoquaux}}},
  \bibinfo{author}{\bibfnamefont{A.}~\bibnamefont{{Gramfort}}},
  \bibnamefont{and} \bibinfo{author}{\bibfnamefont{V.}~\bibnamefont{{Michel}}},
  \bibinfo{journal}{J. Mach. Learn. Res.} \textbf{\bibinfo{volume}{12}},
  \bibinfo{pages}{2825} (\bibinfo{year}{2011}), \eprint{arXiv:1201.0490}.

\bibitem[{\citenamefont{Smith}(2011)}]{fingerprint_region}
\bibinfo{author}{\bibfnamefont{J.~G.} \bibnamefont{Smith}},
  \emph{\bibinfo{title}{Mass spectrometry and infrared spectroscopy}}
  (\bibinfo{publisher}{McGraw-Hill}, \bibinfo{address}{New York},
  \bibinfo{year}{2011}), chap.~\bibinfo{chapter}{13}, p. \bibinfo{pages}{463},
  \bibinfo{edition}{3rd} ed.

\bibitem[{\citenamefont{Hastie et~al.}(2001)\citenamefont{Hastie, Tibshirani,
  and Friedman}}]{elements_of_sl}
\bibinfo{author}{\bibfnamefont{T.}~\bibnamefont{Hastie}},
  \bibinfo{author}{\bibfnamefont{R.}~\bibnamefont{Tibshirani}},
  \bibnamefont{and} \bibinfo{author}{\bibfnamefont{J.}~\bibnamefont{Friedman}},
  \emph{\bibinfo{title}{{The Elements of Statistical Learning}}}, Springer
  Series in Statistics (\bibinfo{publisher}{Springer New York},
  \bibinfo{address}{New York}, \bibinfo{year}{2001}).

\bibitem[{\citenamefont{{Rousseeuw}}(1987)}]{silhouette}
\bibinfo{author}{\bibfnamefont{P.~J.} \bibnamefont{{Rousseeuw}}},
  \bibinfo{journal}{J. Comput. Appl. Math.} \textbf{\bibinfo{volume}{20}},
  \bibinfo{pages}{53} (\bibinfo{year}{1987}).

\bibitem[{\citenamefont{{Li}}(2020)}]{Li2020}
\bibinfo{author}{\bibfnamefont{A.}~\bibnamefont{{Li}}},
  \bibinfo{journal}{NatAs} \textbf{\bibinfo{volume}{4}}, \bibinfo{pages}{339}
  (\bibinfo{year}{2020}).

\bibitem[{\citenamefont{{Kovács} et~al.}(2020)\citenamefont{{Kovács}, {Zhu},
  {Carrete}, {Madsen}, and {Wang}}}]{code}
\bibinfo{author}{\bibfnamefont{P.}~\bibnamefont{{Kovács}}},
  \bibinfo{author}{\bibfnamefont{X.}~\bibnamefont{{Zhu}}},
  \bibinfo{author}{\bibfnamefont{J.}~\bibnamefont{{Carrete}}},
  \bibinfo{author}{\bibfnamefont{G.~K.~H.} \bibnamefont{{Madsen}}},
  \bibnamefont{and} \bibinfo{author}{\bibfnamefont{Z.}~\bibnamefont{{Wang}}}
  (\bibinfo{year}{2020}),
  \urlprefix\url{https://doi.org/10.5281/zenodo.3979217}.

\end{thebibliography}

\end{document}